\documentclass[11pt]{article}
\usepackage{amssymb}
\usepackage{graphics}
\usepackage{epsfig}
\begin{document}
\newcommand{\eepp}{$e^+e^- \to  t \bar t h^0$ }
\newcommand{\eep}{$e^+e^- \to \gamma\gamma \to t \bar t h^0$ }
\newcommand{\ggp}{$\gamma\gamma \to t \bar t h^0$ }

\title {${\cal O}(\alpha_{s})$ QCD and ${\cal O}(\alpha_{ew})$ electroweak
corrections to $t\bar{t}h^0$ production in $\gamma \gamma$
collision \footnote{Supported by National Natural Science
Foundation of China.}} \vspace{3mm}
\author{{ Chen Hui$^{2}$, Ma Wen-Gan$^{1,2}$, Zhang
Ren-You$^{2}$, Zhou Pei-Jun$^{2}$,} \\
{ Hou Hong-Sheng$^{2}$ and Sun Yan-Bin$^{2}$}\\
{\small $^{1}$ CCAST (World Laboratory), P.O.Box 8730, Beijing
100080, P.R.China}\\
{\small $^{2}$ Department of Modern Physics, University of Science and Technology}\\
{\small of China (USTC), Hefei, Anhui 230027, P.R.China}}
\date{}
\maketitle \vskip 12mm

\begin{abstract}

We calculate the ${\cal O}(\alpha_{s})$ QCD and ${\cal
O}(\alpha_{{\rm ew}})$ electroweak one-loop corrections in the
Standard Model framework, to the production of an intermediate
Higgs boson associated with $t\bar{t}$ pair via $\gamma \gamma$
fusion at an electron-positron linear collider (LC). We find the
${\cal O}(\alpha_{s})$ QCD corrections can be larger than the
${\cal O}(\alpha_{{\rm ew}})$ electroweak ones, with the
variations of the Higgs boson mass $m_{h}$ and $e^+e^-$ colliding
energy $\sqrt{s}$. Both corrections may significantly decrease or
increase the Born cross section. The numerical results show that
the relative corrections from QCD to the process \eep may reach
$34.8\%$, when $\sqrt{s}=800$~GeV and $m_h=200$~GeV, while those
from electroweak can be $-13.1\%$, $-15.8\%$ and $-12.0\%$, at
$\sqrt{s} = 800$~GeV, $1$~TeV and $2$~TeV respectively.

\end{abstract}

\vskip 5cm
{\large\bf PACS: 14.65.Ha, 14.80.Bn, 12.15.Lk, 12.38.Bx} \\
{\large\bf Keywords: Associated Higgs boson production, QCD
correction, electroweak correction, photon collider}

\vfill \eject \baselineskip=0.36in
\renewcommand{\theequation}{\arabic{section}.\arabic{equation}}
\renewcommand{\thesection}{\Roman{section}}
\newcommand{\nb}{\nonumber}
\makeatletter      \@addtoreset{equation}{section}
\makeatother
\section{Introduction}

\par
After the discovery of top quark, directly searching for Higgs
boson and studying its property are the main goals of the high
energy colliders. In the electroweak minimal standard model(MSM),
Higgs mechanism generates electroweak symmetry breaking and the
Yukawa coupling terms between Higgs boson and fermions in the
Lagrangian \cite{int higgs1,int higgs2}. In the SM, the Yukawa
term reads ${\cal L}_Y = -\sum_{f}^{} m_f \left( 1+H(x)/v \right)
\bar{\psi}_f \psi_f$. There the coupling strength of the
fermion-Higgs Yukawa coupling $f-\bar{f}-h^0$ is predicted as
$g_{f \bar{f} h} = m_f/v$ at the tree level, where $v =
(\sqrt{2}G_{F})^{-1/2} \simeq 246$~GeV is the vacuum expectation
value of the Higgs field. Since the top quark is the heaviest
fermion, the top quark Yukawa coupling $g_{t \bar{t} h}$ should be
the strongest one among all the fermion-Higgs couplings, e.g.,
$g_{t \bar{t} h}^2 \simeq 0.5$ to be compared for example with
$g_{b \bar{b} h}^2 \simeq 4 \times 10^{-4}$. Therefore, the Higgs
boson production associated with a top-quark pair production
process is particularly important in collider physics for probing
the top-quark Yukawa coupling with the intermediate mass of Higgs
boson.

\par
Recently, LEP2 experiments have provided a lower bound of
$114.4$~GeV for the SM Higgs boson mass at the $95\%$ confidence
level\cite{Lep2}. The future linear colliders (LC) will continue
the work in searching for Higgs boson and studying its property.
There have been already some detailed designs of linear colliders,
such as NLC\cite{NLC}, JLC\cite{JLC}, TESLA\cite{TESLA} and
CLIC\cite{CLIC}. Although the cross section for \eepp process is
small at a LC, about $1$~fb for $\sqrt{s} = 500$~GeV and
$m_h=100$~GeV \cite{ee, ee1, ee2}, it has a distinctive
experimental signature and can potentially be used to measure the
top-quark Yukawa coupling with intermediate Higgs mass at a LC
with very high luminosity.

\par
As we know the apparently the clean signal for light Higgs boson
production associated with a top quark pair is $e^+ e^- \to  t\bar
th^0 \to b\bar b b \bar b W^+W^-$ in both semileptonic and fully
hadronic decay channels. This leads to multi-jet event topologies
involving at lest 6 or more jets in the final state, with $\ge 4$
b-jets and multi-jet invariant mass constraints. Obviously, its
measurement has many difficulties, for example, the tiny signal
with backgrounds about 3 orders of magnitude larger, the
limitations of jet-clustering algorithm in properly reconstructing
multi-jets in the final state, and the degradation of b-tagging
performance due to hard gluon radiation and jet mixing. The
potential backgrounds are from the $t\bar{t}Z^0$ and $t\bar{t}$
productions. The dominant electroweak background is:
\begin{equation}
\gamma \gamma \to t\bar{t}Z^0 \to b\bar{b}W^+W^-Z^0 \to
b\bar{b}b\bar{b}l^{\pm}\nu q\bar{q}'.
\end{equation}
The largest background is from radiative top quark decay:
\begin{equation}
\gamma \gamma \to t\bar{t} \to b\bar{b}W^+W^-g \to
b\bar{b}b\bar{b}l^{\pm}\nu q\bar{q}'.
\end{equation}
Since the b-jets resulting from the gluon splitting are
logarithmically enhanced at low energy, cuts on the jet energy are
efficient at eliminating this background\cite{tth1}. The weak
corrections to \ggp should involve the real emissions of $W$ and
$Z^0$ gauge bosons. These may be other sources of electroweak
backgrounds. Our calculation shows that the cross section of the
real emission of $Z^0$ gauge boson in $t\bar t h^0$ production via
$\gamma \gamma$ collision is under $0.1\%$ of the \ggp process,
and can be neglected in our analysis, while for the real $W$
emission $t\bar t h^0$ production the cross section is in the same
order as that of the \ggp process. In principle, these backgrounds
can easily be removed by using the constraints due to the $W$,
$Z^0$, $t$ and $h^0$ masses with perfect b-tagging and
reconstruction of multi-jets. However, this may not be true in
practice. The experimental situation of the signal detection of
$t\bar t h^0$ production will be even worse, considering the
following two fields. Firstly, the detectors have a finite
coverage in polar angle. Some of these real $W$ bosons can be
emitted so close to the beam direction that part of their decay
products may well escape detection. Secondly, all three particles
in the $t\bar t h^0$ final state are highly unstable, and would
decay in the detector. If $m_h \gtrsim 130~GeV$, it would mostly
produce $b\bar b$ pairs. If above, it would yield four fermions,
either leptons/neutrinos or quarks, principally via $h^0\to
WW^{*}$ decays. Besides, top quark(anti-quark) decays into three
objects, a $b$-quark and the decay products of the $W$,
leptons/neutrinos or quarks. Hence, at tree level in these decays,
we are looking at 8- or 10-fermion final states. In the most
likely case in which the fermions arising from the $W$ decays are
quarks, additional (mainly gluon) radiation will take place, so
that at detector level more than 10 jets may be extracted.
Therefore, the additional real $W$ radiation and $t\bar{t}Z^0$
production entering the detector region could be uneasy to
resolved through their decay products in the data samples of
$t\bar t h^0$ events.

\par
Expected experimental accuracy for determination of the
$t\bar{t}h^0$ coupling in \eepp process has been discussed in many
literatures for specialized linear collider. They show that the
experimental accuracy depends very much on the b-tagging
efficiency. For example, the top quark Yukawa coupling can be
measured to $6-8\%$ accuracy with integral luminosity
$1000~fb^{-1}$ at an $e^+ e^-$ linear collider with
$\sqrt{s}=1~TeV$, assuming $100\%$ efficiency for b-jet tagging
and including statistic but not systematic errors. The accuracy of
the measurement drops to $17-22\%$ if only a $60\%$ efficiency for
b-tagging is achieved \cite{tth1,Baer}. The references
\cite{tth2,tth3,tth4} also stated that the precise determination
of the top-Higgs coupling via the measurement of \eepp process can
reach the accuracy of few percent. Therefore, if we assume the
machine and detector are very efficient and the background
processes can be distinguished substantially, the evaluation of
radiative corrections could be significant for the accurate
experimental measurements of \eepp process. In Ref.\cite{ee1}, S.
Dawson and L. Reina presented the NLO QCD corrections to process
\eepp.  And in references \cite{tthc1,tthc2,tthc3} the electroweak
corrections to the process \eepp are calculated. The
supersymmetric electroweak corrections to process \eepp were
already discussed by X.H. Wu, et al\cite{CSLi}.

\par
An $e^+e^-$ LC can also be designed to operate as a $\gamma\gamma$
collider. This is achieved by using Compton backscattered photons
in the scattering of intense laser photons on the initial $e^+e^-$
beams. The resulting $\gamma -\gamma$ center of mass system (CMS)
energy is peaked at about $0.8\sqrt{s}$ for the appropriate
choices of machine parameters. In $\gamma\gamma$ collision mode at
the high energy peak, we may get approximately the same luminosity
as that of $e^+e^-$ collision. With the new possibility of $\gamma
\gamma$ collisions at electron-positron linear colliders, the
production process \eep offers another approach to probe directly
the top-Higgs coupling in addition to $e^+e^- \to t\bar{t}h^0$ and
$pp(p\bar{p}) \to t\bar{t}h^0$ processes. To detect the Higgs
boson associated with a top-quark pair in high colliding energy,
$\gamma \gamma$ collision has an outstanding advantage over
$e^+e^-$ collision due to its relative larger production rate. The
reason is that at the tree level the \eepp process has a
`s-channel suppression' from the virtual photon and $Z^0$
propagators, especially for the heavy masses of the final
particles. Therefore, we can conclude that the \eep process
provides a better approach than $e^+e^-$ collision to produce
$t\bar{t}h^0$. Similar with the measurement of the process \eepp,
the evaluation of radiative corrections for process \ggp is also
significant for the accurate experimental measurements of top
quark Yukawa coupling.

\par
The Born cross section of \eep process was calculated already in
previous work of Ref.\cite{cheung}. In this paper we neglect the
real gauge boson radiation effects and present the calculations of
the ${\cal O}(\alpha_{s})$ QCD and ${\cal O}(\alpha_{{\rm ew}})$
electroweak corrections to \eep in the SM. We draw a comparison
between our numerical results of Born cross sections of \ggp and
those in Ref.\cite{cheung}. The paper is organized as follow. In
section 2, we present the notations and analytical calculations of
the QCD and the electroweak radiative corrections. The numerical
results and discussions are presented in section 3. Our
conclusions are given in section 4. The numerical comparison of
the Born cross section of process \eep are presented in Appendix.

\section{Analytical Calculation}
\subsection{Calculations of the lowest order of the \ggp subprocess}
\par
\begin{figure}[htb]
\centering
\includegraphics{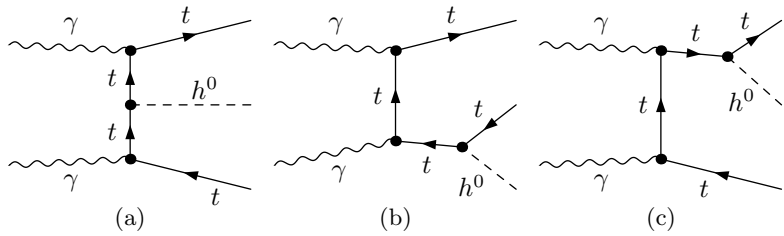}
\caption{The lowest order diagrams for the $\gamma\gamma \to
t\bar{t}h^0$ subprocess.} \label{fig:feyn_born}
\end{figure}

The subprocess \ggp at the lowest level occurs through the u- and
t-channel mechanisms involving Higgs boson bremsstrahlungs
originated from different positions on top-quark lines. The tree
level diagrams are drawn in Fig.\ref{fig:feyn_born}, but the
corresponding diagrams with interchange of the two incoming
photons are not shown.

We denote the subprocess \ggp as
\begin{equation}
\gamma(p_1)+\gamma(p_2) \to t(k_1)+\bar{t}(k_2)+h^0(k_3).
\end{equation}
The four-momenta of incoming electron and positron are denoted as
$p_1$ and $p_2$, respectively, and the four-momenta of outgoing
top-quark, anti-top-quark and Higgs boson are represented as
$k_1$, $k_2$ and $k_3$ correspondingly. All these momenta obey the
on-shell equations $p_1^2=p_2^2=0$, $k_1^2=k_2^2=m_t^2$ and
$k_3^2=m_h^2$.
\par
The amplitudes of the corresponding t-channel Feynman diagrams
(shown in Fig.\ref{fig:feyn_born}(a-c)) of the subprocess $\gamma\gamma \to
t\bar{t}h^0$ are represented by
\begin{eqnarray}
{\cal M}^{(a)}_t=-\frac{e^3
Q_t^2m_t}{2m_W\sin\theta_W}\frac{1}{(k_1-p_1)^2-m_t^2}\frac{1}{(p_2-k_2)^2-m_t^2} ~~~~\nb \\
\times\bar{u}(k_1)\rlap/{\epsilon}(p_1)(\rlap/{k}_1-\rlap/{p}_1+m_t)(\rlap/{p}_2-\rlap/{k}_2+m_t)
\rlap/{\epsilon}(p_2)v(k_2),
\end{eqnarray}
\begin{eqnarray}
{\cal M}^{(b)}_t=-\frac{e^3
Q_t^2m_t}{2m_W\sin\theta_W}\frac{1}{(k_1-p_1)^2-m_t^2}\frac{1}{(k_2+k_3)^2-m_t^2} ~~~~\nb \\
\times\bar{u}(k_1)\rlap/{\epsilon}(p_1)(\rlap/{k}_1-\rlap/{p}_1+m_t)\rlap/{\epsilon}(p_2)
(-\rlap/{k}_2-\rlap/{k}_3+m_t)v(k_2),
\end{eqnarray}
\begin{eqnarray}
{\cal M}^{(c)}_t=-\frac{e^3
Q_t^2m_t}{2m_W\sin\theta_W}\frac{1}{(k_1+k_3)^2-m_t^2}\frac{1}{(p_2-q_2)^2-m_t^2} ~~~~\nb \\
\times\bar{u}(k_1)(\rlap/{k}_1+\rlap/{k}_3+m_t)\rlap/{\epsilon}(p_1)(\rlap/{p}_2-\rlap/{k}_2+m_t)
\rlap/{\epsilon}(p_2)v(k_2),
\end{eqnarray}
where $Q_t=2/3$ and the corresponding amplitudes of the u-channel
Feynman diagrams of the subprocess $\gamma\gamma \to t\bar{t}h^0$
can be obtained by the following interchanges.
\begin{eqnarray}
{\cal M}^{(a)}_u={\cal M}^{(a)}_t(p_1 \leftrightarrow p_2),~~~~
{\cal M}^{(b)}_u={\cal M}^{(b)}_t(p_1 \leftrightarrow p_2),~~~~
{\cal M}^{(c)}_u={\cal M}^{(c)}_t(p_1 \leftrightarrow p_2).
\end{eqnarray}

\par
The total amplitude at the lowest order is the summation of the
above amplitudes.
\begin{equation}
{\cal M}_0=\sum_{i=a,b}^{c} \sum_{j=u}^{t}{\cal M}^{(i)}_{j}.
\end{equation}

\par
Although the previous calculations for the \ggp subprocess and
\eep process at the lowest order were presented by Kingman
Cheung\cite{cheung}, we made the numerical comparison with his
results yet. We calculated the Born cross section of the process
\ggp by using Feynman gauge and unitary gauge to check the gauge
invariance, and adopting FeynArts 3\cite{FA3} and
CompHEP\cite{CompHEP} packages respectively. We found our results
are in good agreement with each other, but not coincident with
Cheung's. The numerical comparisons are presented in Appendix
Table 3.

\vskip 15mm
\subsection{Calculations of the ${\cal O}(\alpha_{s})$ QCD corrections of the \ggp subprocess}

\begin{figure}[htb]
\centering
\includegraphics{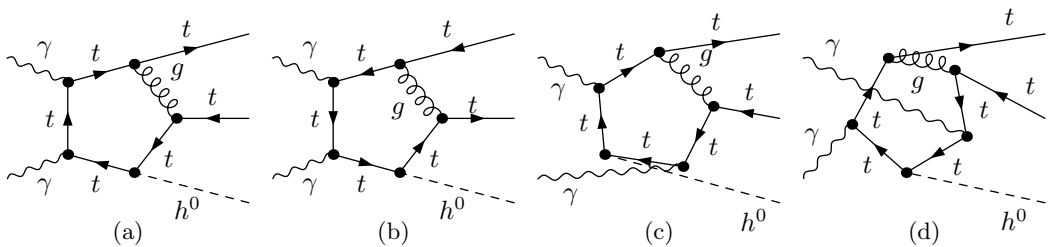}
\caption{ The QCD pentagon diagrams for the $\gamma\gamma \to
t\bar{t}h^0$ subprocess, whose amplitudes include five-point
tensor integrals of rank 4. The corresponding diagrams with
interchange of the two incoming photons are not shown. }
\label{fig:feyn_pen_qcd}
\end{figure}

\par
The ${\cal O}(\alpha_{s})$ QCD one-loop Feynman diagrams of the
subprocess \ggp are generated by ${\it FeynArts}~3$ \cite{FA3}.
There are 84 Feynman diagrams with ${\cal O}(\alpha_{s})$
corrections of the virtual one-loop QCD corrections, which
involves the vertex correction, internal propagator self-energy
correction, box and pentagon diagrams. The Feynman graphs which
generate amplitudes including five-point integrals of rank 4 are
shown in Fig.\ref{fig:feyn_pen_qcd} as a representative selection.
The amplitude of the subprocess \ggp including virtual QCD
corrections to ${\cal O}(\alpha_s)$ can be expressed as
\begin{equation}
{\cal M}_{QCD}={\cal M}_0+\frac{\alpha_s} {4\pi} C_F {\cal
M}_{QCD}^{vir}.
\end{equation}
where $C_F=4/3$, The term $\frac{\alpha_s} {4\pi} C_F {\cal
M}_{QCD}^{vir}$ is the amplitude contributed by the QCD one-loop
Feynman diagrams and the QCD renormalizations of top-quark wave
function, mass and $t-\bar{t}-h^0$ Yukawa coupling. We define the
relevant renormalization constants as
\begin{eqnarray}
 m_{t,0}=m_t+\delta m_{t(g)},~~
t_0^{L}=(1+\frac{1}{2}\delta Z_{t(g)}^L)t^L, ~~
t_0^{R}=(1+\frac{1}{2}\delta Z_{t(g)}^R)t^R,
~~g_{t\bar{t}H}^0=\frac{m_t}{v}+\frac{\delta m_{t(g)}}{v}.
\end{eqnarray}
With the on-mall-shell renormalized condition we get the ${\cal
O}(\alpha_{s})$ QCD contributed parts of the renormalization
constants as
\begin{eqnarray}
\label{counterterm 1}
 \delta m_{t(g)} &=& \frac{ m_t}{2}
        \tilde{Re} \left ( \Sigma_{t(g)}^{L}(m_{t}^{2}) +
               \Sigma_{t(g)}^{R}(m_{t}^{2}) +
                2 \Sigma_{t(g)}^{S}(m_{t}^{2})  \right ),
\end{eqnarray}
\begin{eqnarray}
\label{counterterm 2} \delta Z_{t(g)}^{L} &=& -
\tilde{Re}\Sigma_{t(g)}^{L}(m_{t}^{2})
    - m_{t}^2 \frac{\partial}{\partial p^2}
        \tilde{Re}\left [ \Sigma_{t(g)}^{L}(p^2) +
      \Sigma_{t(g)}^{R}(p^2) + 2\Sigma_{t(g)}^{S}(p^2) \right ] |_{p^2=m_{t}^2},
\end{eqnarray}
\begin{eqnarray}
\label{counterterm 3}
\delta Z_{t(g)}^{R} &=&
-\tilde{Re}\Sigma_{t(g)}^{R}(m_{t}^{2}) -
    m_{t}^2 \frac{\partial}{\partial p^2}
        \tilde{Re}\left [ \Sigma_{t(g)}^{L}(p^2) +
                \Sigma_{t(g)}^{R}(p^2) +
                2 \Sigma_{t(g)}^{S}(p^2)  \right ] |_{p^2=m_{t}^2},
\end{eqnarray}
$\tilde{Re}$ takes the real part of the loop integrals appearing
in the self energies only. Here we define the renormalized
top-quark irreducible two-point function as
\begin{eqnarray}
\hat{\Gamma}_{t}(p^2)=i[\rlap/p P_L \hat{\Sigma}_{t}^{L}(p^2) +
\rlap/p P_R \hat{\Sigma}_{t}^{R}(p^2)+ m_{t}
\hat{\Sigma}_{t}^{S}(p^2)] \delta_{\alpha \beta}
\end{eqnarray}
where $\alpha$ and $\beta$ are the color indices of the top quarks
on the two sides of the self-energy diagram,
$P_{L,R}=(1\mp\gamma_5)/2$. The unrenormalized top-quark self
energy parts contributed by ${\cal O}(\alpha_s)$ QCD read
\begin{eqnarray}
\Sigma_{t(g)}^{L}(p^2)=\Sigma_{t(g)}^{R}(p^2)=\frac{g_s^2}{6
\pi^2}\left(-1+2B_0[p,0,m_t]+2B_1[p,0,m_t]\right)
\end{eqnarray}
and
\begin{eqnarray}
\Sigma_{t(g)}^{S}(p^2)=\frac{g_s^2}{3
\pi^2}\left(1-2B_0[p,0,m_t]\right)
\end{eqnarray}

The corresponding contribution part to the cross section at ${\cal
O}(\alpha_{s})$ order can be written as
\begin{eqnarray}
\Delta \hat{\sigma}^{QCD}_{vir} = \hat{\sigma}_{0}
\hat{\delta}^{QCD}_{vir} = \frac{\alpha_s C_F}{8\pi |\vec{p}_1|
\sqrt{\hat{s}}} \int {\rm d} \Phi_3 \overline{\sum_{{\rm spin}}}
{\rm Re} \left( {\cal M}_{0}^{\dag} {\cal M}_{QCD}^{vir} \right),
\end{eqnarray}
where ${\rm d} \Phi_3$ is the three-body phase space element. The
bar over summation recalls averaging over initial spins.

\par
The virtual QCD corrections contain both ultraviolet (UV) and
infrared (IR) divergences in general. To regularize the UV
divergences in loop integrals, we adopt the dimensional
regularization in which the dimensions of spinor and spacetime
manifolds are extended to $D = 4 - 2 \epsilon$. We have verified
the cancellation of the UV both analytically and numerically. Then
we get a UV finite amplitude including ${\cal O}(\alpha_{s})$
virtual radiative corrections.

\par
The IR divergence in the ${\cal M}_{QCD}^{vir}$ of the process
\ggp is originated from virtual gluon corrections. It can be
exactly cancelled by including the real gluon bremsstrahlung
corrections to this subprocess in the soft gluon limit. The real
gluon emission process is denoted as
\begin{eqnarray}
\label{real gluon emission}
 \gamma(p_1)+\gamma(p_2) \rightarrow t(k_1)+\bar{t}(k_2)+h^0(k_3)+g(k),
\end{eqnarray}
where the real gluon radiates from the internal or external
top(anti-top) quark line, and can be classified into two parts
which behave soft and hard natures respectively. In order to
isolate the soft gluon emission singularity in the real gluon
emission process, we adopt the general phase-space-slicing method
\cite{PSS}, in which the bremsstrahlung phase space is divided
into singular and non-singular regions. The cross section of the
real gluon emission process ($\ref{real gluon emission}$) is
decomposed into soft and hard terms
\begin{equation}
\Delta \hat{\sigma}_{real}^{QCD}=\Delta \hat{\sigma}_{
soft}^{QCD}+\Delta \hat{\sigma}_{hard}^{QCD}=
\hat{\sigma}_0(\hat{\delta}_{soft}^{QCD}+\hat{\delta}_{hard}^{QCD}).
\end{equation}
We adopt the soft gluon approximation method with a cut $\Delta E$
in numerical calculations for the soft emission corrections
($k^0<\Delta E$). With this approach the real gluon radiation from
internal color lines does not lead to IR singularities and can be
neglected in this approach. We find the contribution of the soft
gluon emission process is \cite{COMS,Velt}
\begin{eqnarray}
\label{approsoft QCD} {\rm d} \Delta \hat{\sigma}_{{\rm
soft}}^{QCD} = -{\rm d} \hat{\sigma}_0 \frac{\alpha_{s}C_F}{2
\pi^2}
 \int_{|\vec{k}| \leq \Delta E}\frac{{\rm d}^3 k}{2 k^0} \left[
 \frac{k_1}{k_1\cdot k}-\frac{k_2}{k_2\cdot k} \right]^2,
\end{eqnarray}
in which $\Delta E$ is the energy cutoff of the soft gluon and
$k^0 \leq \Delta E \ll \sqrt{s}$. $k^0 = \sqrt{|\vec{k}|^2+m_g^2}$
is the gluon energy. Here we have introduced a small gluon mass
$m_g$ to regulate the infrared divergences occurring in the soft
emission. The integral over the soft gluon phase space have been
implemented, the analytical result of the soft gluon corrections
to $\gamma\gamma \to t\bar{t}h^0g$ is presented as\cite{ee1}
\begin{eqnarray*}
\label{soft g emmision} \left(\frac{{\rm
d}\Delta\hat{\sigma}^{QCD}_{soft}}{{\rm d}x_h}\right)
&=&\left(\frac{{\rm d}\hat{\sigma}_{0}}{{\rm d}x_h}\right )
\frac{\alpha_sC_F}{2\pi} \left\{ -2 \log\left( \frac{4\Delta
E^2}{m_g^2}\right)\left[ 1-\frac{xk_1 \cdot
k_2}{m_t^2(x^2-1)}\log(x^2) \right]-\frac{k_1^0}{|\vec{k}_1|}
\log\left(\frac{k_1^0-|\vec{k}_1|}{k_1^0+|\vec{k}_1|}\right)  \right.  \\
&-& \left. \frac{k_2^0}{|\vec{k}_2|}
\log\left(\frac{k_2^0-|\vec{k}_2|}{k_2^0+|\vec{k}_2|}\right)+\frac{4xk_1
\cdot k_2}{m_t^2(x^2-1)} \left[ \frac{1}{4} \log^2\left(
\frac{u^0-|\vec{u}|}{u^0+|\vec{u}|}\right)+{\rm Li}_2
\left( 1-\frac{u^0+|\vec{u}|}{v} \right) \right. \right. \\
&+& \left. \left. {\rm Li}_2 \left(
1-\frac{u^0-|\vec{u}|}{v}\right) \right]_{u=xk_2}^{u=xk_1}
\right\},
\end{eqnarray*}
where
\begin{equation}
v=\frac{m_t^2(x^2-1)}{2(xk_1^0-k_2^0)},
\end{equation}
and $x$ is the solution of
\begin{equation}
m_t^2(x^2+1)-2xk_1\cdot k_2=0,
\end{equation}
which should satisfy the constraint of
\begin{equation}
\frac{xk_1^0-k_2^0}{k_2^0} > 0.
\end{equation}

\par
We checked numerically the cancellation of IR divergencies and
verified that the contribution of these soft gluon bremsstrahlung
corrections leads to an IR finite cross section which is
independent of the infinitesimal gluon mass $m_g$. The hard gluon
emission cross section $\Delta \hat{\sigma}^{QCD}_{{\rm hard}}$
for $E_g>\Delta E$, is calculated numerically by using Monte Carlo
method. The statistic error is controlled under $0.3\%$.

\par
Finally the UV and IR finite total cross section of the subprocess
\ggp  including the ${\cal O}(\alpha_{s})$ QCD corrections reads
\begin{equation}\label{cs}
\hat{\sigma}^{QCD}  = \hat{\sigma}_0 + \Delta \hat{\sigma}^{QCD} =
\hat{\sigma}_0 +  \Delta \hat{\sigma}_{{\rm vir}}^{QCD} +
 \Delta \hat{\sigma}_{{\rm real}}^{QCD} = \hat{\sigma}_0(1 +
\hat{\delta}^{QCD}),
\end{equation}
where $\hat{\delta}^{QCD} = \hat{\delta}_{{\rm vir}}^{QCD} +
\hat{\delta}_{{\rm soft}}^{QCD} + \hat{\delta}_{{\rm hard}}^{QCD}
$ is the QCD relative correction of order ${\cal O}(\alpha_s)$.

\vskip 10mm
\subsection{The calculation of the ${\cal O}(\alpha_{{\rm ew}})$ one-loop corrections to the \ggp subprocess}
\par

\par
\begin{figure}[htb]
\centering
\includegraphics{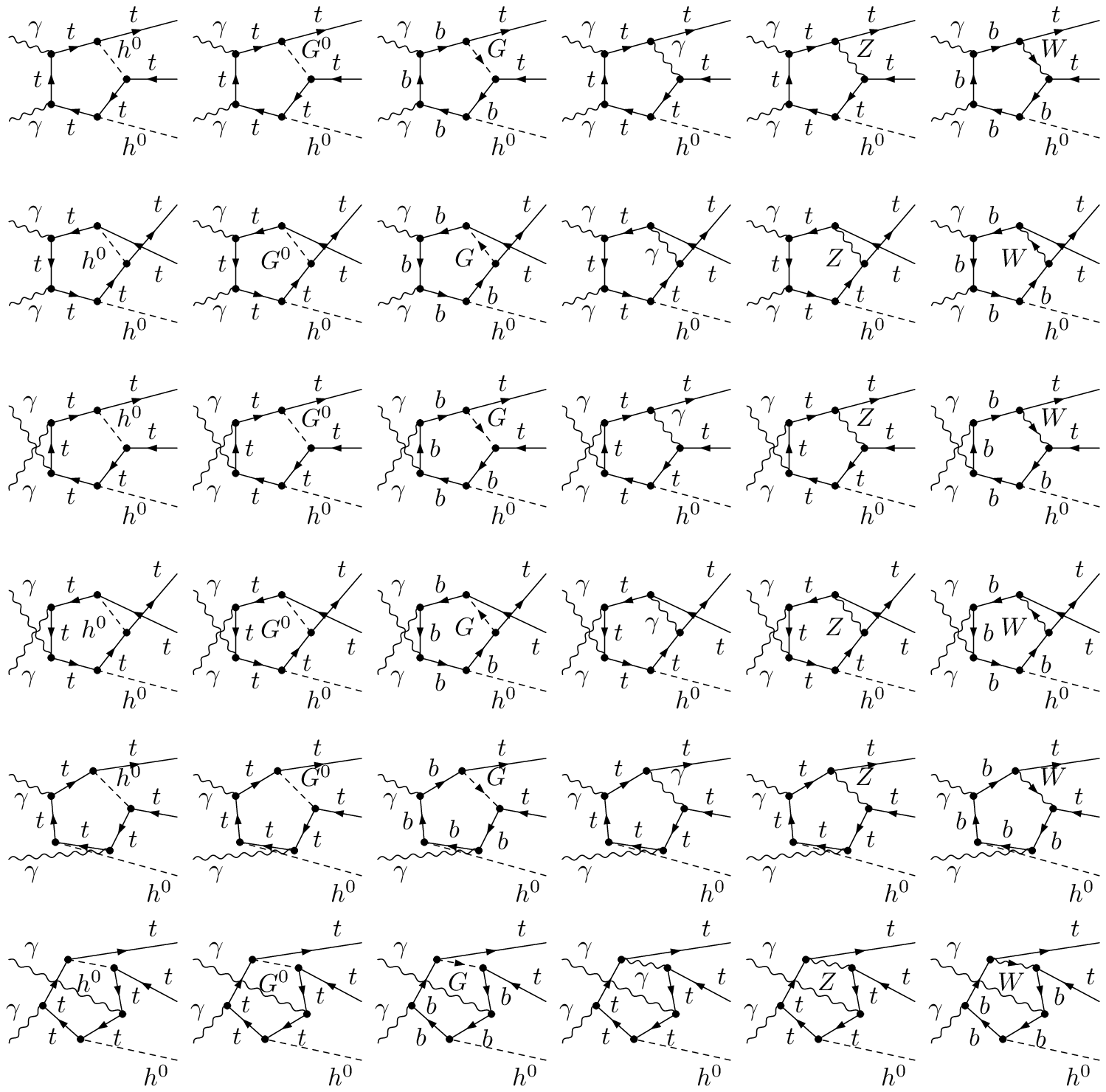}
\caption{ The five-point pentagon electroweak one-loop diagrams
for the $\gamma\gamma \to t\bar{t}h^0$ subprocess, whose
corresponding amplitudes include five-point tensor integrals of
rank 4. } \label{fig:feyn_pen}
\end{figure}

In addition to the QCD corrections, we also calculate the ${\cal
O}(\alpha_{{\rm ew}})$ one-loop electroweak corrections to the
subprocess \ggp. We use again the package {\it FeynArts}
3\cite{FA3} to generate the electroweak one-loop Feynman diagrams
and the relevant amplitudes of the subprocess \ggp automatically.
The electroweak one-loop Fenyman diagrams can be classified into
self-energy, triangle, box and pentagon diagrams. The pentagon
diagrams, whose corresponding amplitudes include five-point tensor
integrals of rank 4, are depicted in Fig.\ref{fig:feyn_pen} as a
representative selection. In our electroweak correction
calculation we use also the t'Hooft-Feynman gauge and adopt the
definitions of one-loop integral functions of Ref.\cite{s14}. The
one-loop level virtual electroweak corrections to $\gamma \gamma
\rightarrow t\bar{t} h^0$ can be expressed as
\begin{eqnarray}
\Delta \hat{\sigma}^{EW}_{{\rm vir}} = \hat{\sigma}_{0}
\hat{\delta}^{EW}_{{\rm vir}} = \frac{N_c}{2 |\vec{p}_1|
\sqrt{\hat{s}}} \int {\rm d} \Phi_3 \overline{\sum_{{\rm spin}}}
{\rm Re} \left( {\cal M}^{\dag}_{0} {\cal M}^{EW}_{{\rm vir}}
\right),
\end{eqnarray}
where $\vec{p}_1$ is the c.m.s. three-momentum of one of the
incoming photons, ${\rm d} \Phi_3$ is the three-body phase space
element, and the bar over summation recalls averaging over initial
spins. ${\cal M}^{EW}_{{\rm vir}}$ is the amplitude of the
electroweak one-loop Feynman diagrams, including self-energy,
vertex, box, pentagon and counterterm diagrams.

\par
Analogously to the case of the QCD correction calculations shown
in last subsection, we adopt the dimensional regularization scheme
to regularize the UV divergences in loop integrals. We assume that
there is no quark mixing, the KM-matrix is identity matrix and use
the complete on-mass-shell (COMS) renormalization scheme
\cite{COMS}, in which the electric charge of electron $e$ and the
physical masses $m_W$, $m_Z$, $m_h$, $m_t$ et al., are chosen to
be the renormalized parameters. The relevant field renormalization
constants are defined as \cite{COMS}
\begin{eqnarray}
e_0=(1+\delta Z_e)e,~~~ m_{h,0}^2=m_h^2+\delta m_h^2, ~~~
m_{t,0}=m_t+\delta m_t,  ~~~m_{W,0}^2=m_W^2+\delta m_W^2, ~~~ \nb \\
m_{Z,0}^2=m_Z^2+\delta m_Z^2, ~~~A_0=\frac{1}{2}\delta
Z_{AZ}Z+(1+\frac{1}{2}\delta Z_{AA})A,~~~~
h_0=(1+\frac{1}{2}\delta Z_h)h,  ~~  \nb \\
t_0^{L}=(1+\frac{1}{2}\delta Z^L)t^L,  ~~~~
t_0^{R}=(1+\frac{1}{2}\delta Z^R)t^R.
\end{eqnarray}
With the on-mass-shell conditions, we can obtain the renormalized
constants expressed as
\begin{eqnarray}
\delta m_W^2 = \tilde{Re} \Sigma_T^W(m_W^2),~~~\delta m_Z^2 = Re
\Sigma_T^{ZZ}(m_Z^2),  ~~~ \delta Z_{AA}=- \frac{\partial
\Sigma_T^{AA}(p^2)}{\partial p^2}|_{p^2=0},~~~ \nb \\
\delta Z_{ZZ}= - Re \frac{\partial \Sigma_T^{ZZ}(p^2)}{\partial
p^2}|_{p^2=m_Z^2}, ~~~ \delta
Z_{ZA}=2\frac{\Sigma_T^{ZA}(0)}{m_Z^2}, ~~~
\delta Z_{AZ}= - 2 Re
\frac{\Sigma_T^{AZ}(m_Z^2)}{m_Z^2}, ~~~\nb \\
\delta m_h=Re\Sigma^h(m_h^2),~~~ \delta Z_h=-Re \frac{\partial
\Sigma^h(p^2)}{\partial p^2}|_{p^2=m_h^2}.
\end{eqnarray}
The renormalization constants of the wave function and mass of
top-quark can be evaluated from Eqs.(\ref{counterterm
1})-(\ref{counterterm 3}) upon replacing the QCD top-quark
self-energies($\Sigma_{t(g)}^{L}$, $\Sigma_{t(g)}^{R}$ and
$\Sigma_{t(g)}^{S}$) by the electroweak corresponding ones
($\Sigma_{t}^{L}$, $\Sigma_{t}^{R}$ and $\Sigma_{t}^{S}$),
respectively. And the explicit expressions of the electroweak self
energies in the SM concerned in our calculations can be found in
the Appendix B of Ref.\cite{COMS}. The UV divergence appearing
from the one-loop diagrams is cancelled by the contributions of
the counterterm diagrams in our calculations. We have verified
both analytically and numerically that the final cross sections
including ${\cal O}(\alpha_{{\rm ew}})$ virtual radiative
corrections and the corresponding counterterm contributions are UV
finite. The further verification of the correctness for the
one-loop calculation is made by probing the gauge independence of
cross section via changing the value of $\xi$ in $R_{\xi}$ gauge,
and the results are coincident with each other very well.

\par
Analogous to the calculation of the QCD corrections, the IR
divergence in the subprocess \ggp originates from virtual photonic
corrections is cancelled by the real photonic bremsstrahlung
corrections in the soft photon limit. We use also the general
phase-space-slicing method \cite{PSS} and divide the phase space
into singular and non-singular regions. Then the cross section of
the real photon emission subprocess ($\gamma \gamma \to t \bar{t}
h^0 \gamma$) is decomposed into soft and hard parts.
\begin{equation}
\Delta \hat{\sigma}_{{\rm real}}=\Delta \hat{\sigma}_{{\rm
soft}}+\Delta \hat{\sigma}_{{\rm hard}}= \hat{\sigma}_0(
\hat{\delta}^{\gamma}_{{\rm soft}}+ \hat{\delta}^{\gamma}_{{\rm
hard}}).
\end{equation}
By using the soft photon($k^0<\Delta E$) approximation, we get the
contribution of the soft photon emission subprocess expressed as
\cite{COMS,Velt}
\begin{eqnarray}
\label{approsoft} {\rm d} \Delta \hat{\sigma}^{\gamma}_{{\rm
soft}} = -{\rm d} \hat{\sigma}_0 \frac{\alpha_{{\rm ew}}Q_t^2}{2
\pi^2}
 \int_{|\vec{k}| \leq \Delta E}\frac{{\rm d}^3 k}{2 k_0} \left[
 \frac{k_1}{k_1\cdot k}-\frac{k_2}{k_2\cdot k} \right]^2,
\end{eqnarray}
in which $\Delta E$ is the energy cutoff of the soft photon and
$k^0 \leq \Delta E \ll \sqrt{\hat{s}}$,  $Q_t=2/3$ is the electric
charges of top quark, $k^0 = \sqrt{|\vec{k}|^2+m_{\gamma}^2}$ is
the photon energy. Therefore, after the integration over the soft
photon phase space, we obtain the analytical result of the soft
corrections to \ggp. Actually, the real photonic emission
correction can be deduced from the real gluon emission corrections
upon replacing the factor $C_F \alpha_s$ by $Q^2 \alpha_{ew}$. The
cancellation of IR divergencies is verified and the results of the
cross section show the independence on the infinitesimal photon
mass $m_{\gamma}$ in our calculation.
\par
Since sometimes the QED radiative contributions can be quite
large, the investigation of the genuine weak corrections
quantitatively would help us to understand the origination of the
large electroweak corrections. The QED correction to the
subprocess \ggp is gauge invariant and comprises three parts: (1)
photonic virtual radiations of final and internal top-quarks, (2)
real photonic radiations from final and internal top-quarks, (3)
the interference of final and internal real photon radiations.
Actually, the total ${\cal O}(\alpha_{{\rm ew}})$ order QED
corrections for the subprocess \ggp, can be obtained numerically
from the results of the ${\cal O}(\alpha_{s})$ order QCD
corrections as presented in section II.2 through multiplying a
factor of $\frac {\alpha_{ew} Q_t^2}{ \alpha_{s}(\mu) C_F}$. Then
the genuine weak corrections can be evaluated by subtracting the
QED corrections from the ${\cal O}(\alpha_{{\rm ew}})$ QED
corrections. We define the genuine weak relative correction as,
\begin{eqnarray}
\label{genuine weak corre} \hat{\delta}_w =
\hat{\delta}-\hat{\delta}^{{QED}}=\hat{\delta}-\hat{\delta}^{{QED}}_{vir}-
\hat{\delta}^{{QED}}_{soft}-\hat{\delta}^{{QED}}_{hard}.
\end{eqnarray}
where $\hat{\delta}^{{QED}}_{vir}$ is the relative correction
contributed by the QED one-loop diagrams including virtual photon
exchange and the corresponding parts of the counter terms.
$\hat{\delta}^{{QED}}_{soft}$ and $\hat{\delta}^{{QED}}_{hard}$
are the relative corrections of the soft and hard real photon
emissions, respectively.

\vskip 10mm
\subsection{Calculations of process \eep }
\par
By using the laser back-scattering technique on electron beam, an
$e^+e^-$ LC which has c.m.s. energy of hundred GeV to several TeV
can be transformed to be a photon collider \cite{Com1,Com2,Com3}.
By integrating over the photon luminosity in an $e^+e^-$ linear
collider, the total cross section of the process \eep can be
obtained in the form as
\begin{eqnarray}
\label{integration} \sigma(s)= \int_{E_{0}/ \sqrt{s}} ^{x_{max}} d
z \frac{d {\cal L}_{\gamma\gamma}}{d z} \hat{\sigma}(\gamma\gamma
\to t \bar t h \hskip 3mm
 at \hskip 3mm  \hat{s}=z^{2} s)
\end{eqnarray}
where $E_{0}=2m_{t}+m_{h}$, and $\sqrt{s}$($\sqrt{\hat{s}}$) being
the $e^{+}e^{-}$($\gamma\gamma$) c.m.s. energy. $\frac{d\cal
L_{\gamma\gamma}}{d z}$ is the distribution function of photon
luminosity, which is defined as:
\begin{eqnarray}
\frac{d{\cal L}_{\gamma\gamma}}{dz}=2z\int_{z^2/x_{max}}^{x_{max}}
 \frac{dx}{x} F_{\gamma/e}(x)F_{\gamma/e}(z^2/x)
\end{eqnarray}
For the initial unpolarized electrons and laser photon beams, the
energy spectrum of the back scattered photon is given by
\cite{phospec}
\begin{eqnarray}
\label{structure}
F_{\gamma/e}=\frac{1}{D(\xi)}\left[1-x+\frac{1}{1-x}-
\frac{4x}{\xi(1-x)}+\frac{4x^{2}}{\xi^{2}(1-x)^2}\right]
\end{eqnarray}
where $x=2 \omega/\sqrt{s}$ is the fraction of the energy of the
incident electron carried by the back-scattered photon, the
maximum fraction of energy carried by the back-scattered photon is
$x_{max}=2 \omega_{max}/\sqrt{s}=\xi/(1+\xi)$, and
\begin{eqnarray}
D(\xi)=(1-\frac{4}{\xi}-\frac{8}{{\xi}^2})\ln{(1+\xi)}+\frac{1}{2}+
  \frac{8}{\xi}-\frac{1}{2{(1+\xi)}^2},
\end{eqnarray}
\begin{eqnarray}
  \xi=\frac{2\sqrt{s} \omega_0}{{m_e}^2}.
\end{eqnarray}
$m_{e}$ and $\sqrt{s}/2$ are the mass and energy of the electron,
$\omega_0$ is the laser-photon energy. In our evaluation, we
choose $\omega_0$ such that it maximizes the backscattered photon
energy without spoiling the luminosity through $e^{+}e^{-}$ pair
creation. Then we have ${\xi}=2(1+\sqrt{2})$, $x_{max}\simeq
0.83$, and $D(\xi) \approx 1.84$, as used in Ref.\cite{photon
para}.

\section{Numerical results and discussions}

\par
For the numerical calculation we use the following input
parameters \cite{pdg}
\begin{table}[htb]
  \begin{center}
    \begin{tabular}{lcllcllcl}
$\alpha_{{\rm ew}}(0)^{-1}$ &=& $137.03599976$,&
$m_W$ & = & $80.423$~GeV,& $m_Z$ & = & $91.1876$~GeV, \\
$m_e$ & = & $0.510998902$~MeV, & $m_\mu$ &=& $105.658357$~MeV, &
$m_\tau$ &=& $1.77699$~GeV, \\
$m_u$ & = & $66$~MeV, & $m_c$ & = & $1.2$~GeV, &
$m_t$ & = & $174.3$~GeV, \\
$m_d$ & = & $66$~MeV, & $m_s$ & = & $150$~MeV, & $m_b$ & = &
$4.3$~GeV, \\
$\alpha_s(m_Z^2)$ & = & $0.117186$.
\end{tabular}
  \end{center}
\end{table}

where we use the effective values of the light quark masses ($m_u$
and $m_d$) which can reproduce the hadron contribution to the
shift in the fine structure constant $\alpha_{{\rm ew}}(m_Z^2)$
\cite{jeger}, and take the pole mass of top quark($m_t=174.3~GeV$)
to determine the $t\bar{t}h^0$ coupling strength. The QCD
renormalization scale $\mu$ is taken to be $(2 m_t+m_h)/2$ and the
running of the strong coupling $\alpha_s(\mu^2)$ is evaluated at
the two-loop level($\overline{MS}$ scheme) with five active
flavors.
\par
\begin{figure}[htb]
\centering
\includegraphics{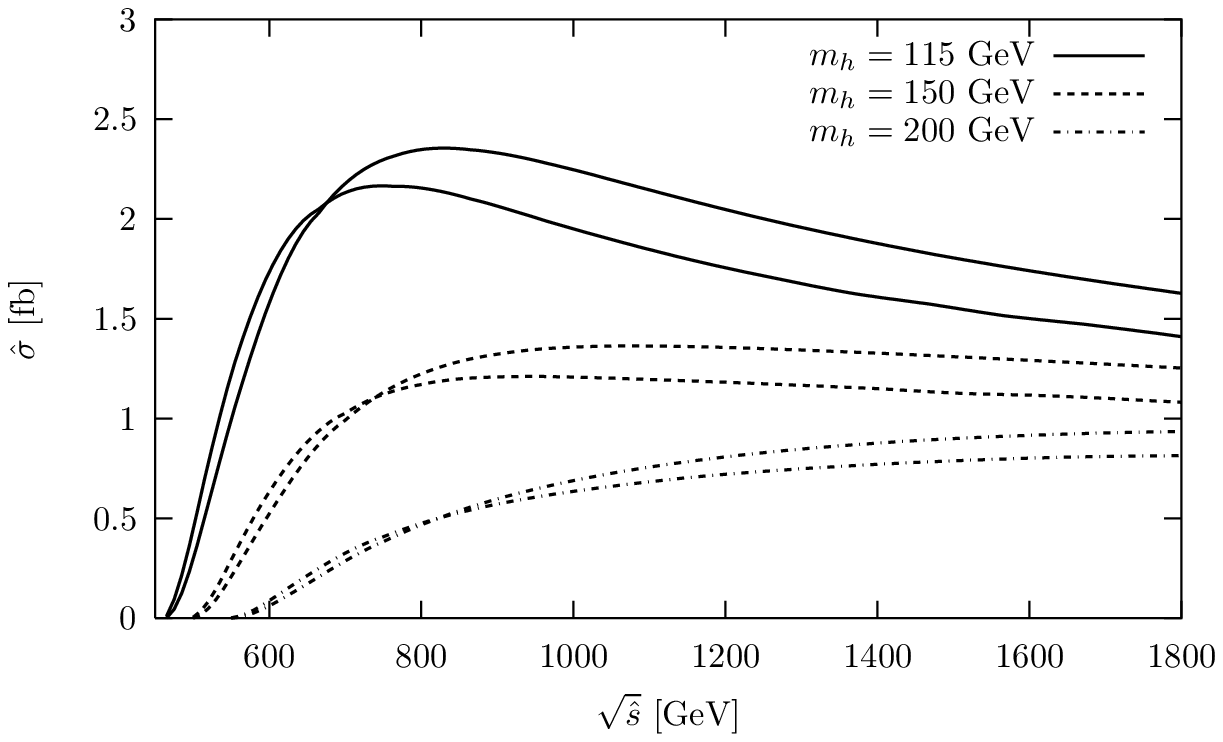}
\caption{The Born and one-loop QCD corrected cross sections for
the \ggp subprocess as the functions of c.m.s. energy
($\sqrt{\hat{s}}$) with $m_h=115$, $150$, $200$~GeV, respectively.
For each line type, the upper curve(in the energy region
$\sqrt{\hat{s}}>1~TeV$) is for the Born cross section and the
lower one for the one-loop QCD corrected cross section.}
\label{fig:gg_s_cs_qcd}
\end{figure}
\begin{figure}[htb]
\centering
\includegraphics{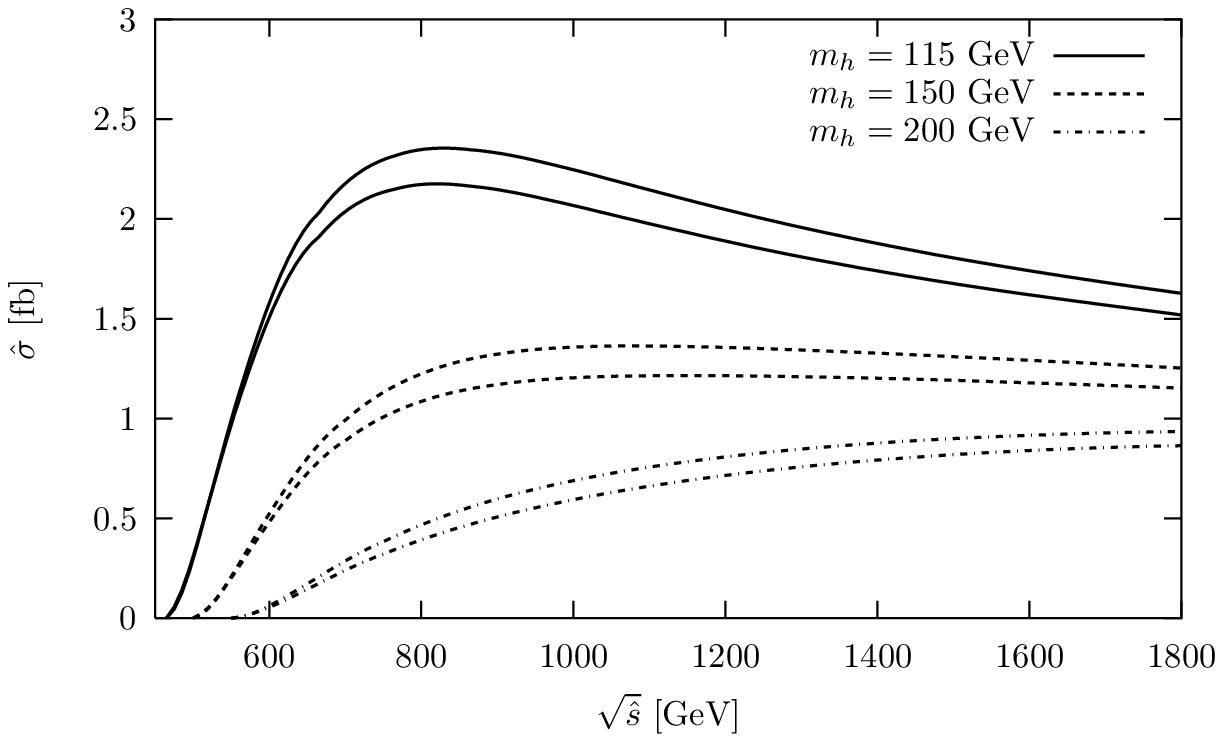}
\caption{The Born and one-loop electroweak corrected cross
sections for the \ggp subprocess as the functions of c.m.s. energy
($\sqrt{\hat{s}}$) with $m_h=115$, $150$, $200$~GeV. For each line
type, the upper curve is for the Born cross section and the lower
one for the one-loop electroweak corrected cross section.}
\label{fig:gg_s_cs}
\end{figure}

The numerical results of the cross sections with QCD and one-loop
electroweak radiative corrections for the subprocess \ggp are
plotted in Fig.\ref{fig:gg_s_cs_qcd} and Fig.\ref{fig:gg_s_cs},
respectively. The full, dashed, and dash-dotted curves correspond
to the cases with $m_h=115$, $150$ and $200$ GeV, correspondingly,
and $\gamma \gamma$ colliding energy $\sqrt{\hat{s}}$ runs from
the value little larger than the threshold ($2 m_t+m_h$) to $1.8$
TeV. For each line type there are two curves, the upper curve (in
the region $\sqrt{\hat{s}}>1~TeV$) is for the Born cross section
and the lower one for the QCD corrected cross section. As
indicated in Fig.\ref{fig:gg_s_cs_qcd}, the QCD corrections can
increase (when $\sqrt{\hat{s}} < 650~GeV$) or decrease the cross
sections of subprocess \ggp(when $\sqrt{\hat{s}} > 900~GeV$),
while Fig.\ref{fig:gg_s_cs} shows that the one-loop electroweak
radiative corrections always reduce the Born cross sections in the
plotted energy range of $\sqrt{\hat{s}}$. The curves in both
Fig.\ref{fig:gg_s_cs_qcd} and Fig.\ref{fig:gg_s_cs} show that all
the Born, QCD and electroweak corrected cross sections decrease
with the increment of the mass of Higgs boson $m_h$. The curves
for $m_h=115$~GeV increase rapidly to their corresponding maximal
cross section values, when the $\gamma \gamma$ colliding energy
$\sqrt{\hat{s}}$ goes from the threshold value to the
corresponding position of peak. The curves for $m_h=150$~GeV have
platforms when $\sqrt{\hat{s}}$ are larger than $1000$~GeV, and
for $m_h=200$~GeV both Born and one-loop corrected cross sections
increase slowly in the whole plotted range of $\sqrt{\hat{s}}$.

\par
\begin{figure}
\centering
\includegraphics{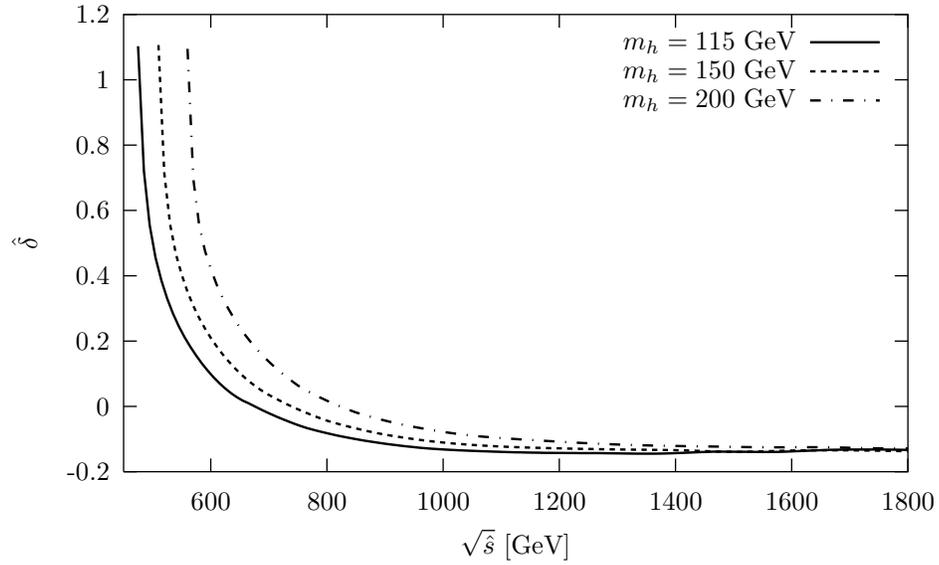}
\caption{ The QCD one-loop relative corrections as the functions
of c.m.s. energy ($\sqrt{\hat{s}}$) for the $\gamma\gamma \to
t\bar{t}h^0$ subprocess with $m_h=115$, $150$, $200$~GeV,
respectively. } \label{fig:gg_s_rel_qcd}
\end{figure}
\begin{figure}
\centering
\includegraphics{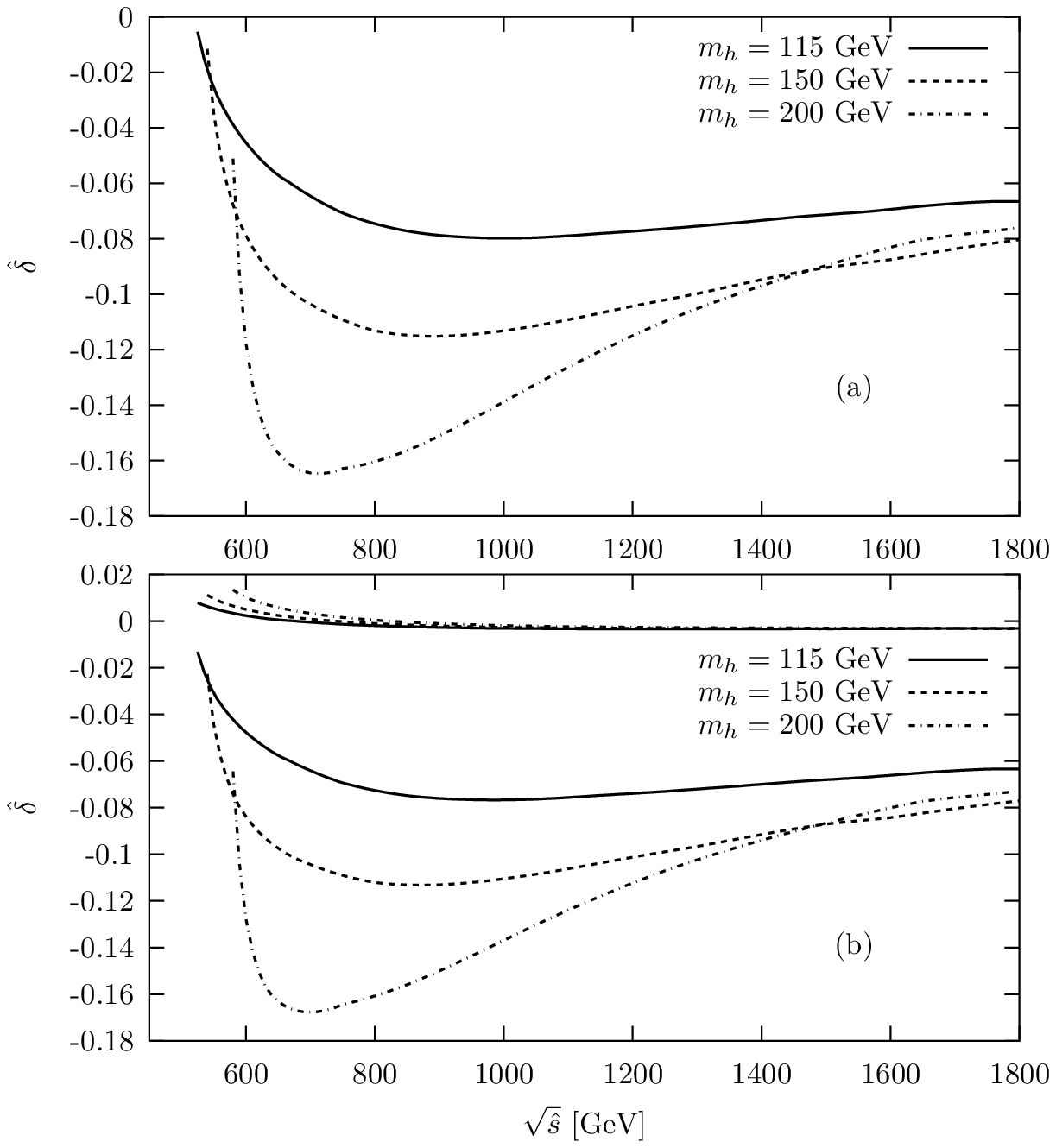}
\caption{(a)The electroweak one-loop relative corrections as the
functions of c.m.s. energy ($\sqrt{\hat{s}}$) with $m_h=115$,
$150$, $200$~GeV for the $\gamma\gamma \to t\bar{t}h^0$
subprocess. (b)The QED and weak one-loop relative corrections as
the functions of c.m.s. energy ($\sqrt{\hat{s}}$) with $m_h=115$,
$150$, $200$~GeV for the $\gamma\gamma \to t\bar{t}h^0$
subprocess. For each line type, the upper curve is for the QED
corrected cross section and the lower one for the weak corrected
cross section.}\label{fig:gg_s_rel}
\end{figure}

We define the QCD relative corrections to the subprocess \ggp and
parent process \eep as
\begin{eqnarray}
\hat{\delta}^{QCD}=\frac{\hat{\sigma}^{QCD}-\hat{\sigma}_0}{\hat{\sigma}_0},~~~~~~~~~
\delta^{QCD}=\frac{\sigma^{QCD}-\sigma_0}{\sigma_0},
\end{eqnarray}
respectively, and the electroweak relative corrections to the
subprocess \ggp and process \eep as
\begin{eqnarray}
\hat{\delta}=\frac{\hat{\sigma}-\hat{\sigma}_0}{\hat{\sigma}_0},~~~~~~~~~~~~~~~
\delta=\frac{\sigma-\sigma_0}{\sigma_0}, ~~~~\nb \\
\end{eqnarray}
correspongdingly. The ${\cal O}(\alpha_{s})$ QCD relative
corrections and ${\cal O}(\alpha_{{\rm ew}})$ electroweak relative
corrections to the cross sections for \ggp subprocess,
corresponding to Fig.\ref{fig:gg_s_cs_qcd} and
Fig.\ref{fig:gg_s_cs}, are depicted in Fig.\ref{fig:gg_s_rel_qcd}
and Fig.\ref{fig:gg_s_rel}(a-b), respectively.
We can read from Fig.\ref{fig:gg_s_rel_qcd} that the QCD relative
corrections to the subprocess \ggp decrease from $111\%$ to
$-13.8\%$ when the c.m.s. energy $\sqrt{\hat{s}}$ increases from
the threshold energy to $1.8$~TeV. From Fig.7(a) we can see that
in the plotted colliding energy range, the ${\cal O}(\alpha_{{\rm
ew}})$ order electroweak relative corrections to subprocess \ggp
can reach $-7.98\%$ and $-16.5\%$ for $m_h=115$~GeV and $200$~GeV,
respectively. The maximal electroweak absolute relative
corrections to the cross sections $|\hat{\delta}|_{max}$ and the
corresponding $\sqrt{\hat{s}}$ positions for subprocess \ggp with
$m_h=115$, $130$, $150$, $170$, $200$~GeV are listed in Table
\ref{tab:gg_s_max}. In Fig.7(b) the ${\cal O}(\alpha_{{\rm ew}})$
order QED and genuine weak relative corrections are depicted,
respectively. It shows that the ${\cal O}(\alpha_{{\rm ew}})$ QED
relative corrections are very small comparing with the genuine
weak relative corrections, and can only reach $1.28\%$ for the
curve of $m_h=200~GeV$ at the position of $\sqrt{\hat{s}} \sim
580~GeV$.

\begin{table}[p]
  \begin{center}
    \begin{tabular}{|c|c|c|}\hline
       $m_h$ (GeV) & $\sqrt{\hat{s}}$ (GeV)
        & $|\hat{\delta}|_{max} (\%)$ \\\hline
        115 & 1000 & 7.98\\
    130 & 950 & 9.15\\
    150 & 890 & 11.5\\
    170 & 820 & 14.0\\
    200 & 710 & 16.5\\
      \hline
    \end{tabular}
\caption{The maximum electroweak absolute relative corrections and
the corresponding colliding energy $\sqrt{\hat{s}}$ positions for
the \ggp subprocess with $m_h=115$, $130$, $150$, $170$,
$200$~GeV, respectively. } \label{tab:gg_s_max}
  \end{center}
\end{table}
\par
\begin{figure}
\centering
\includegraphics{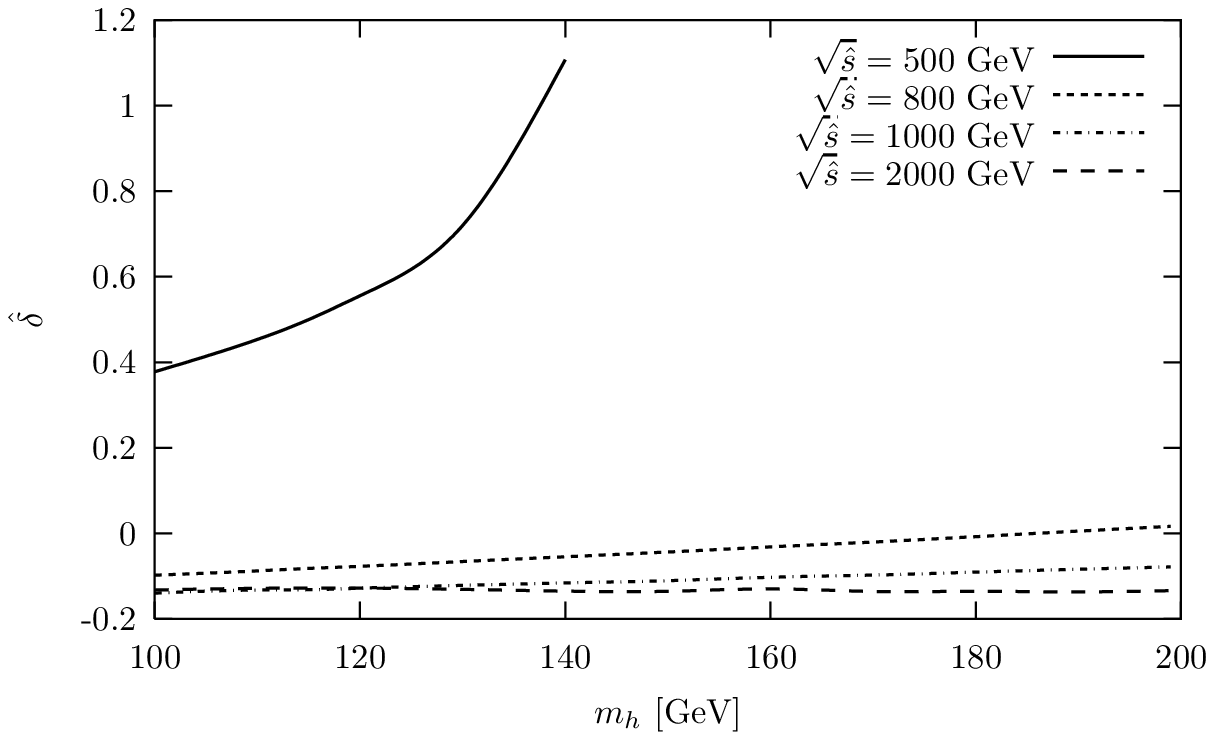}
\caption{ The QCD relative corrections to the cross sections of
\ggp subprocess as the functions of the mass of Higgs boson
($m_h$) with $\sqrt{\hat{s}}=500$, $800$, $1000$, $2000$~GeV,
respectively.} \label{fig:gg_h_rel_qcd}
\end{figure}
\begin{figure}
\centering
\includegraphics{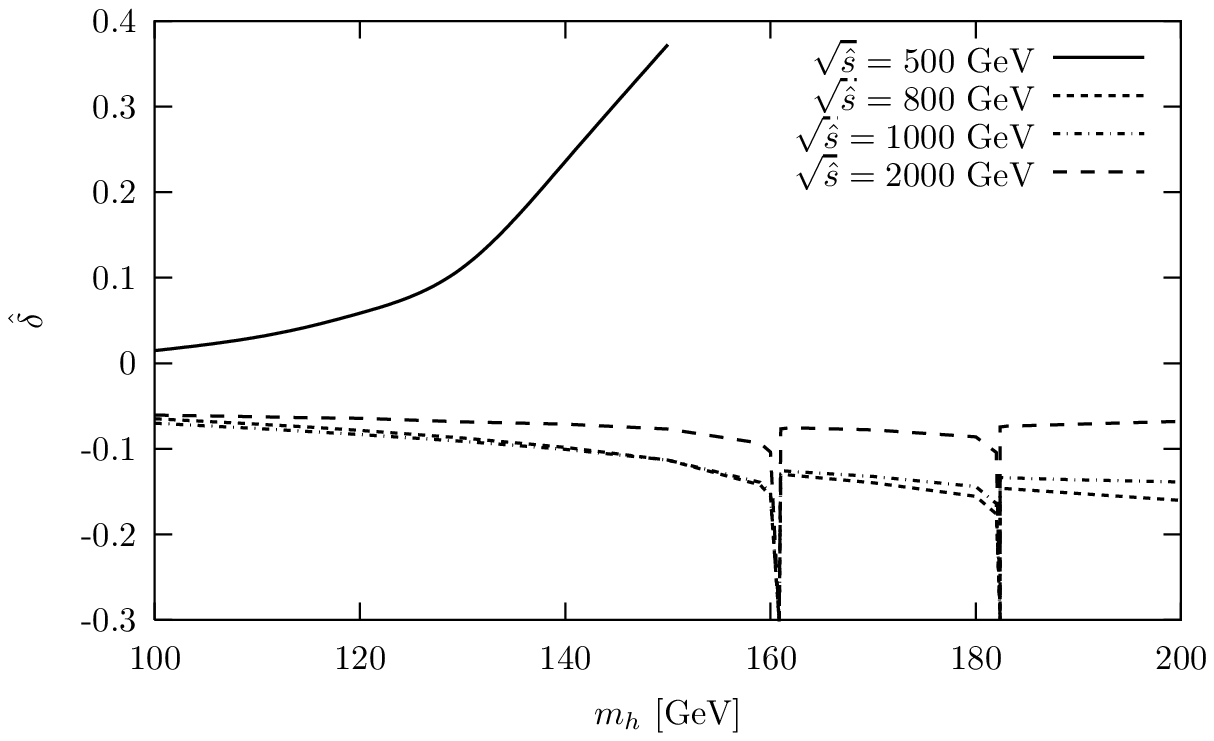}
\caption{ The electroweak relative corrections to the cross
sections of \ggp subprocess as the functions of the mass of Higgs
boson ($m_h$) with $\sqrt{\hat{s}}=500$, $800$, $1000$,
$2000$~GeV, respectively.} \label{fig:gg_h_rel}
\end{figure}

We also depicted the QCD and electroweak relative corrections to
the cross sections of subprocess \ggp as the functions of the mass
of Higgs boson $m_h$ with $\sqrt{\hat{s}}=500$, $800$, $1000$,
$2000$ GeV in Fig.\ref{fig:gg_h_rel_qcd} and
Fig.\ref{fig:gg_h_rel}, respectively. Both curves for
$\sqrt{\hat{s}}=500$~GeV in Fig.\ref{fig:gg_h_rel_qcd} and
Fig.\ref{fig:gg_h_rel} are truncated before the position of $m_h
\sim 150$~GeV, because the channel \ggp cannot be opened when
$\sqrt{\hat{s}} < m_h+2 m_t$. On the curves for
$\sqrt{\hat{s}}=800$, $1000$, $2000$~GeV in
Fig.\ref{fig:gg_h_rel}, there exist two resonance peaks on each
curve at the positions $m_h \sim 2m_Z$ and $m_h \sim 2m_W$. Since
we didn't consider the widths of $W^{\pm}$ and $Z^0$ gauge bosons
in loop calculation, the genuine one-loop weak corrections in the
vicinities of the thresholds at $m_h=2m_W$ and $m_h=2m_Z$ shown in
Fig.\ref{fig:gg_h_rel} are untrustworthy. In
Fig.\ref{fig:gg_h_rel_qcd}, the QCD relative corrections for
$\sqrt{\hat{s}}=500$~GeV are rather large, and can reach the value
larger than $110\%$ when the $m_h=140~GeV$ and
$\sqrt{\hat{s}}=500~GeV$, while the QCD relative corrections are
in the range $1.74\% \sim -14.0\%$, when $\sqrt{\hat{s}}=800$,
$1000$, $2000$~GeV. Fig.\ref{fig:gg_h_rel} shows that the curves
of the electroweak relative corrections for
$\sqrt{\hat{s}}=800$~GeV and $1000$~GeV go down from $-6.48\%$ to
$-16.0\%$ and from $-7.01\%$ to $-13.9\%$ respectively, when $m_h$
varies from $100$~GeV to $200$~GeV. While the electroweak relative
corrections for $\sqrt{\hat{s}}=2000$~GeV are relative stable(
about minus a few percent) except in the Higgs mass regions
satisfying resonance conditions.

\begin{figure}
\centering
\includegraphics{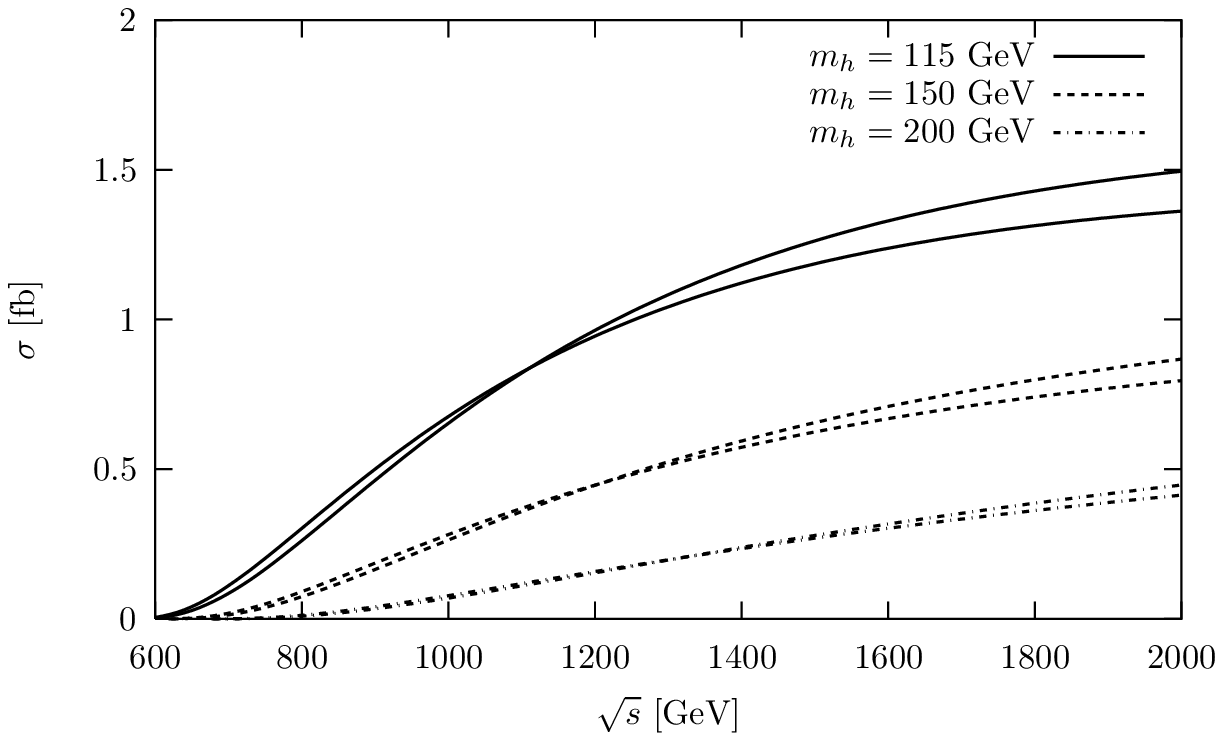}
\caption{ The Born and one-loop QCD corrected cross sections for
the \eep process as the functions of c.m.s. energy ($\sqrt{s}$)
with $m_h=115$, $150$, $200$~GeV, respectively. For each line
type, the upper curve(in the energy region $\sqrt{s} > 1.4~TeV$)
is for the Born cross section and the lower one presents the
one-loop corrected cross section. } \label{fig:ee_s_cs_qcd}
\end{figure}
\begin{figure}
\centering
\includegraphics{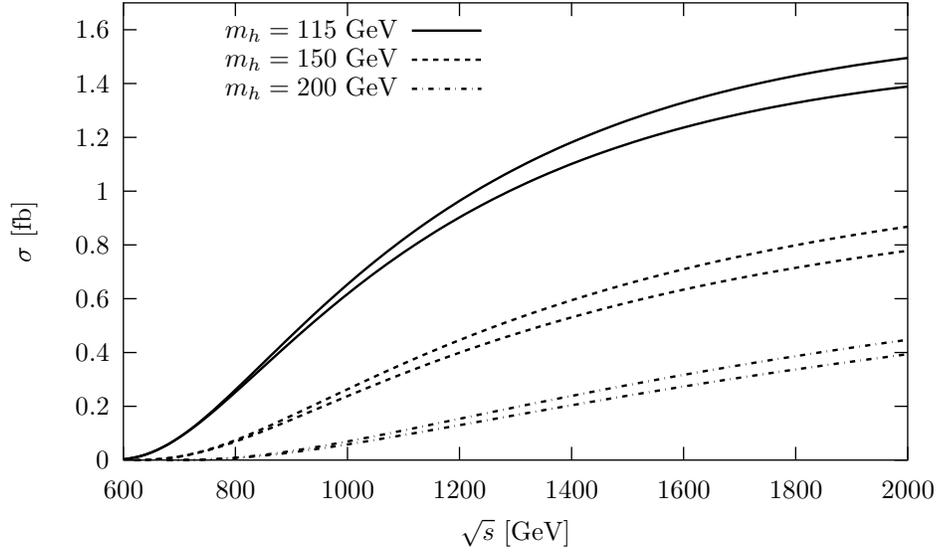}
\caption{ The Born and one-loop electroweak corrected cross
sections for the \eep process as the functions of c.m.s. energy
($\sqrt{s}$) with $m_h=115$, $150$, $200$~GeV, respectively. For
each line type, the upper curve is for the Born cross section and
the lower one presents the one-loop corrected cross section.}
\label{fig:ee_s_cs}
\end{figure}
\begin{figure}
\centering
\includegraphics{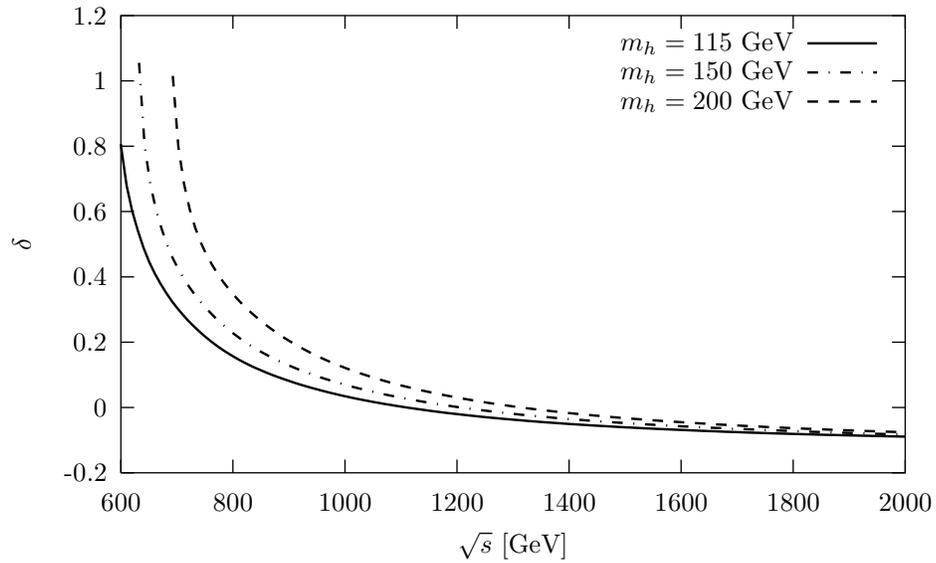}
\caption{The QCD relative corrections to the \eep process as the
functions of c.m.s. energy ($\sqrt{s}$) with $m_h=115$, $150$,
$200$~GeV, respectively.} \label{fig:ee_s_rel_qcd}
\end{figure}
\begin{figure}
\centering
\includegraphics{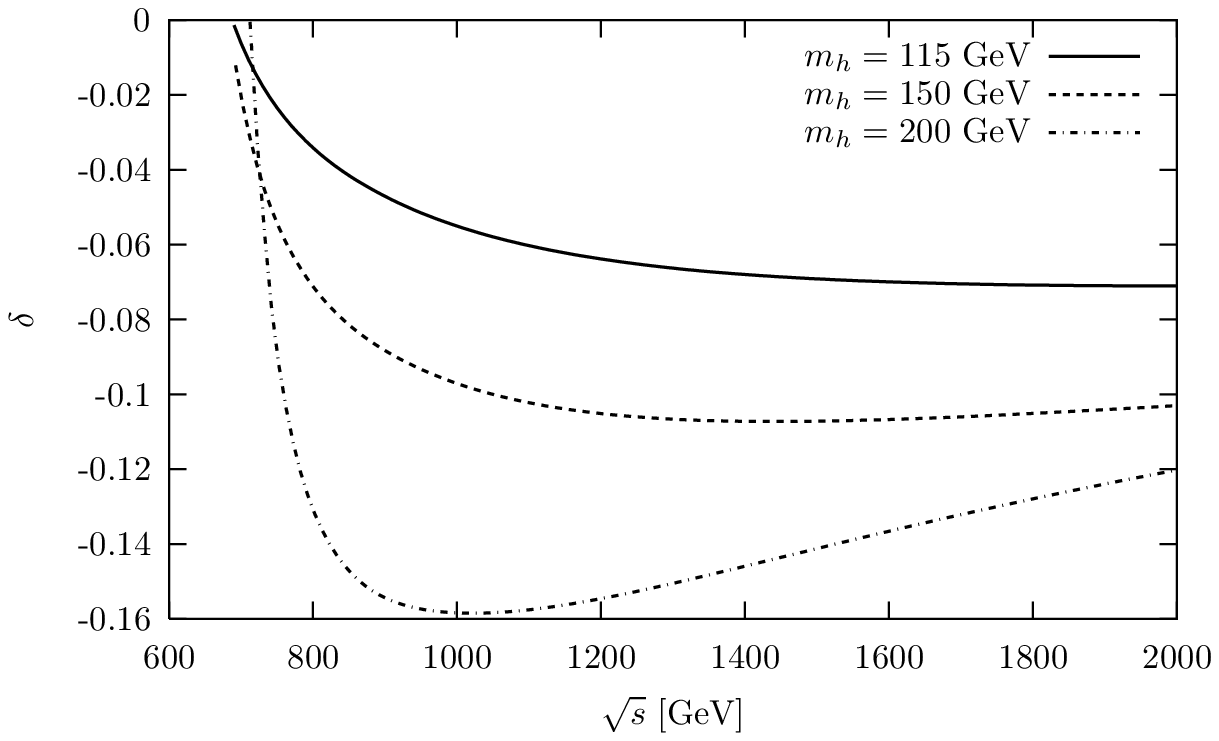}
\caption{The electroweak relative corrections to the \eep process
as the functions of c.m.s. energy ($\sqrt{s}$) with $m_h=115$,
$150$, $200$~GeV, respectively.} \label{fig:ee_s_rel}
\end{figure}

\par
Fig.\ref{fig:ee_s_cs_qcd} and Fig.\ref{fig:ee_s_cs} show the cross
sections including ${\cal O}(\alpha_{s})$ QCD and ${\cal
O}(\alpha_{ew})$ radiative corrections and for \eep process versus
the $e^+e^-$ colliding energy $\sqrt{s}$, respectively. Both
figures demonstrate that the Born and radiative corrected cross
sections increase with the increment of $\sqrt{s}$, and the ${\cal
O}(\alpha_{s})$ QCD radiative corrections for different values of
$m_h$ can reduce or increase the Born cross sections, but the
${\cal O}(\alpha_{{\rm ew}})$ electroweak corrections reduce the
Born cross section only in the plotted range of electron-positron
c.m.s. energy $\sqrt{s}$. The QCD and electroweak relative
corrections for the \eep parent process corresponding to Fig.10
and Fig.11, are depicted in Fig.12 and Fig.13, respectively. We
can see that the QCD relative correction can be very large near
the threshold colliding energy, while the electroweak relative
corrections can reach $-15.9\%$ for $m_h=200$~GeV in the vicinity
of $\sqrt{s} \sim 1~TeV$. The maximum absolute electroweak
relative corrections for different Higgs boson mass values in the
plotted colliding c.m.s. energy range can be read out from this
figure and are listed in Table \ref{tab:ee_s_max}.
\begin{table}[htb]
  \begin{center}
    \begin{tabular}{|c|c|c|}\hline
       $m_h$ (GeV) & $\sqrt{s}$ (GeV)
        & $|\delta|_{max} (\%)$ \\\hline
        115 & 2000 & 7.10\\
    130 & 1700 & 8.28\\
    150 & 1435 & 10.7\\
    170 & 1220 & 13.4\\
    200 & 1020 & 15.9\\
      \hline
    \end{tabular}

\caption{The maximum absolute relative corrections and the
corresponding colliding energy $\sqrt{s}$ positions for the \eep
process with $m_h=115$, $130$, $150$, $170$, $200~$GeV,
respectively.  } \label{tab:ee_s_max}
\end{center}
\end{table}

\begin{figure}
\centering
\includegraphics{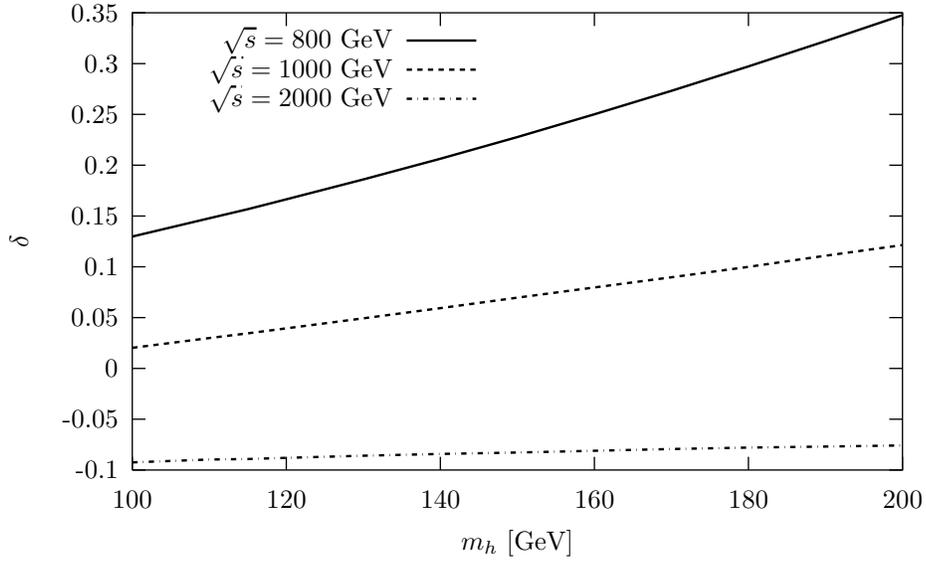}
\caption{The QCD relative corrections to the \eep process as the
functions of Higgs-boson mass ($m_h$) with $\sqrt{s}=800$, $1000$,
$2000$~GeV, respectively.} \label{fig:ee_h_rel_qcd}
\end{figure}
\begin{figure}
\centering
\includegraphics{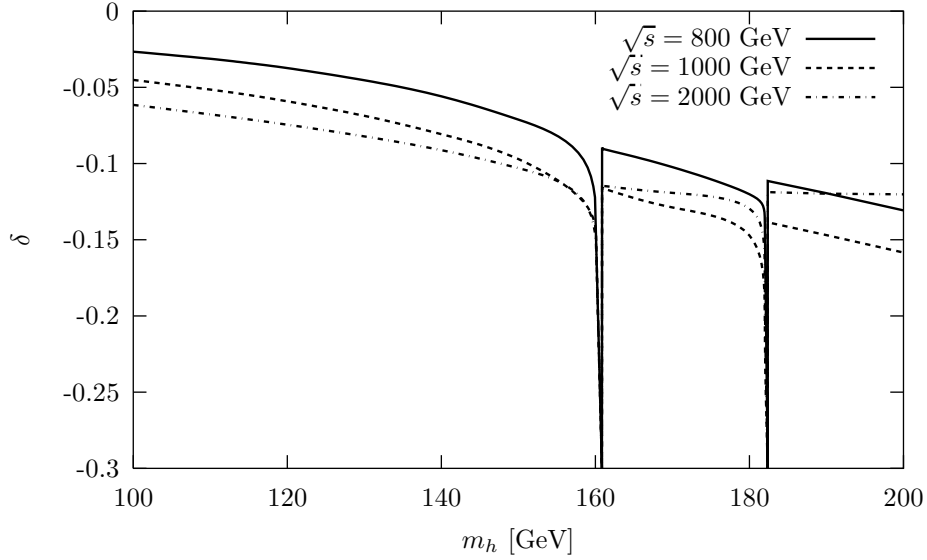}
\caption{The electroweak relative corrections to the \eep process
as the functions of Higgs-boson mass ($m_h$) with $\sqrt{s}=800$,
$1000$, $2000$~GeV, respectively.} \label{fig:ee_h_rel}
\end{figure}
\begin{figure}
\centering
\includegraphics{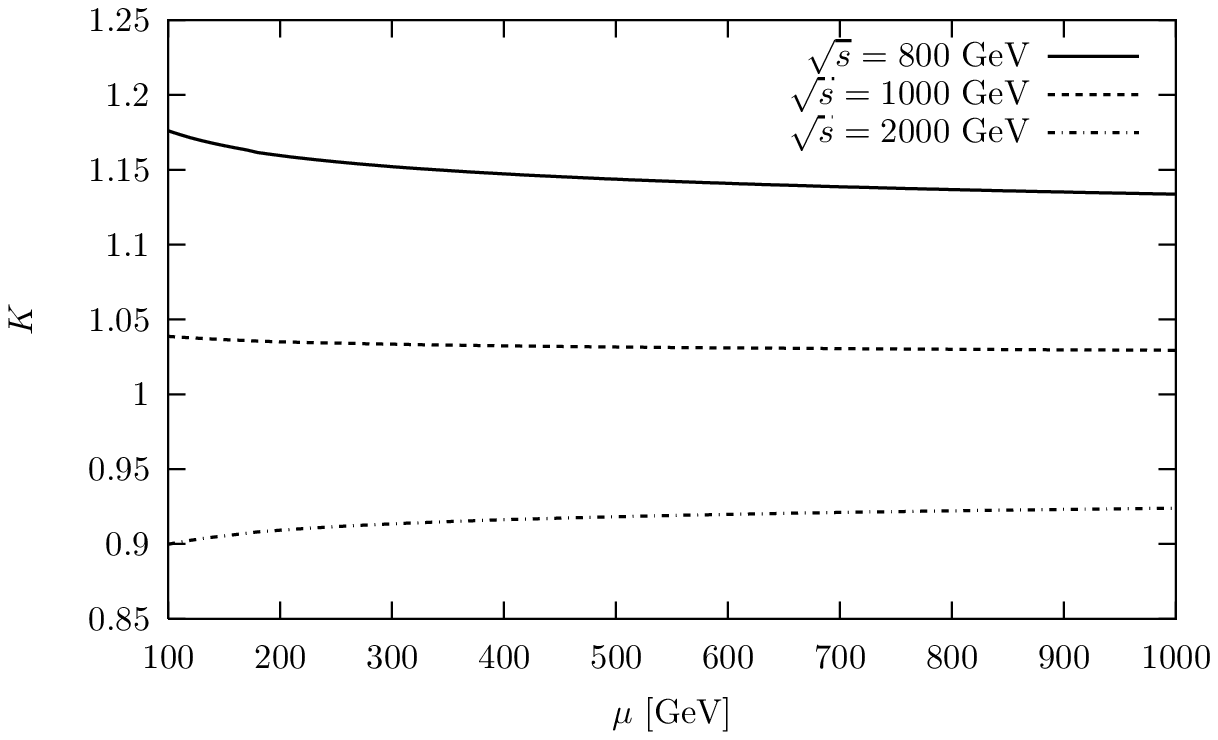}
\caption{ The QCD relative corrections to the \eep process as the
function of the QCD renormalization scale $\mu$ with $m_h =
115~GeV$ and $\sqrt{s} = 800,~1000,~2000~GeV$.}
\label{fig:ee_mu_rel_qcd}
\end{figure}

\par
The QCD and electroweak relative corrections to the cross section
of the process \eep as the functions of the Higgs boson mass $m_h$
are depicted in Fig.\ref{fig:ee_h_rel_qcd} and
Fig.\ref{fig:ee_h_rel} with $\sqrt{s}=800$, $1000$, $2000$~GeV,
respectively. From Fig.\ref{fig:ee_h_rel_qcd} we can see that the
QCD one-loop relative correction for $\sqrt{s}=800$~GeV can reach
$34.8\%$ at the position of $m_h=200$~GeV. In
Fig.\ref{fig:ee_h_rel} two resonance peaks appear again on each
curve at the positions of $m_h \sim 2m_Z$ and $m_h \sim 2m_W$
because of the resonance effects, and the correction values in
these vicinities are untrustworthy. For the curves of
$m_h=800$~GeV and $m_h=1000$~GeV, the relative correction decrease
from $-2.66\%$ and $-4.51\%$ to $-13.1\%$ and $-15.9\%$,
respectively, when $m_h$ varies from $100$~GeV to $200$~GeV. The
relative correction for $\sqrt{s}=2000$~GeV varies in the range of
[$-6.15\%,-12.0\%$] when $m_h$ goes from $100$~GeV to $200$~GeV,
except in the $h^0$ mass regions satisfying the resonance effect
conditions.

\par
Scale dependence of the $K$-factors
($=\frac{\sigma^{QCD}}{\sigma_{0}}$) of the total cross sections
for the process \eep at LC ($m_h=115~GeV$) is plotted in
Fig.\ref{fig:ee_mu_rel_qcd}. The full-line, dashed-line and
dash-dotted-line correspond to $\sqrt{s}=800~GeV$, $1~TeV$ and
$2~TeV$, respectively. We can see from the figure that when the
scale $\mu$ goes from $100~GeV$ to $1~TeV$, the QCD $K$-factors
vary from 1.176, 1.039, 0.900 to 1.134, 1.029, 0.924 for
$\sqrt{s}=800~GeV$, $1000~GeV$, $2000~GeV$, respectively. We can
conclude that the theoretical uncertainty of the QCD correction at
${\cal O}(\alpha_s)$ level due to the variation of energy scale
$\mu$, is under $4\%$ for $\sqrt{s}=800,~1000,~2000~GeV$, when
$m_h=115~GeV$ and energy scale $\mu$ is in the range of $100~GeV$
to $1~TeV$.

\section{ Summary}
\par
In this paper we calculate the ${\cal O}(\alpha_{s})$ QCD and
${\cal O}(\alpha_{{\rm ew}})$ one-loop electroweak radiative
corrections to the \eep process with Higgs boson in intermediate
mass region at an $e^+e^-$ linear collider (LC) in the SM. We
investigate the dependence of the QCD and electroweak radiative
corrections to both subprocess \ggp and parent process \eep on the
Higgs boson mass $m_{h}$ and colliding energy $\sqrt{\hat{s}}$
($\sqrt{s}$), and find that the QCD corrections can either
increase or decrease the Born cross section, while the electroweak
corrections always decrease the Born cross section of the \eep
parent process and \ggp subprocess in the Higgs boson mass range
$115$~GeV$ < m_h < 200$~GeV. We also notice that the ${\cal
O}(\alpha_{s})$ QCD corrections to process \eep can be larger than
the ${\cal O}(\alpha_{{\rm ew}})$ corrections depending on the
Higgs boson mass $m_{h}$ and $e^+e^-$ colliding energy $\sqrt{s}$.
Both kinds of corrections may significantly decrease or increase
the Born cross sections. The numerical results show that the
${\cal O}(\alpha_{s})$ QCD relative corrections to the process
\eep can reach $34.8\%$ when $\sqrt{s}=800$~GeV and $m_h=200$~GeV,
while the ${\cal O}(\alpha_{{\rm ew}})$ electroweak relative
corrections to the Born cross sections can reach $-13.1\%$,
$-15.8\%$ and $-12.0\%$ at $\sqrt{s} = 800$~GeV, $1$~TeV and
$2$~TeV, respectively.

\vskip 5mm
\noindent{\large\bf Acknowledgments:} This work was
supported in part by the National Natural Science Foundation of
China and a grant from the University of Science and Technology of
China.

\vskip 10mm
\begin{flushleft} {\bf Appendix} \end{flushleft}
In Appendix we list the numerical comparison of the cross sections
at the tree level for the process \eep. In order to check our
calculation we use two independently developed packages {\it
FeynArts} 3\cite{FA3} and CompHEP\cite{CompHEP} to evaluate the
cross sections. The results of ours and Kingman Cheung's
\cite{cheung} are presented in Table \ref{table-2}. It is clear
that our cross sections at tree level are not coincident with
Cheung's results.

\begin{table}
$$\begin{array}{c@{\quad}c@{\quad}c@{\quad}c@{\quad}c@{\quad}c}
\hline m_t~ [{\rm GeV}] & m_h~ [{\rm GeV}] & \sqrt{s}~ [{\rm GeV}]
&\sigma~ [{\rm fb}]~ ({\rm Ref.\cite{cheung}}) & \sigma~ [{\rm fb}]~ ({\rm FeynArts}) &
\sigma~ [{\rm fb}]~ ({\rm CompHEP}) \\
\hline
120 & 60  & 500  & 0.45 &0.391(0) & 0.391(8) \\
    &     & 1000 & 2.6  &2.18(7) & 2.19(1) \\
    &     & 2000 & 2.8  &2.39(1) & 2.39(1) \\
\hline
150 & 60  & 1000 & 3.2  &2.74(1) & 2.74(5) \\
    &     & 2000 & 4.1  &3.42(1) & 3.42(2) \\
    & 140 & 1000 & 0.36 &0.311(8)& 0.311(6) \\
    &     & 2000 & 0.95 &0.805(9)& 0.805(6) \\
\hline
180 & 140 & 1000 & 0.40 &0.341(3)& 0.341(5) \\
    &     & 2000 & 1.2  &1.05(5)& 1.05(5) \\
\hline
\end{array}$$
\caption{The numerical comparison of the cross sections of the
process \eep at tree-level with the results in Ref.\cite{cheung}
by using FeynArts 3 and CompHEP packages.} \label{table-2}
\end{table}

\vskip 10mm
\begin{flushleft} {\bf Figure Captions} \end{flushleft}
\par
{\bf Figure \ref{fig:feyn_born}} The lowest order diagrams for the
$\gamma\gamma \to t\bar{t}h^0$ subprocess..

\par
{\bf Figure \ref{fig:feyn_pen_qcd}} The QCD pentagon diagrams for
the $\gamma\gamma \to t\bar{t}h^0$ subprocess, whose amplitudes
include five-point tensor integrals of rank 4. The corresponding
diagrams with interchange of the two incoming photons are not
shown.

\par
{\bf Figure \ref{fig:feyn_pen}} The five-point pentagon
electroweak one-loop diagrams for the $\gamma\gamma \to
t\bar{t}h^0$ subprocess, whose corresponding amplitudes include
five-point tensor integrals of rank 4.

\par
{\bf Figure \ref{fig:gg_s_cs_qcd}} The Born and one-loop QCD
corrected cross sections for the \ggp subprocess as the functions
of c.m.s. energy ($\sqrt{\hat{s}}$) with $m_h=115$, $150$,
$200$~GeV, respectively. For each line type, the upper curve (in
the energy region $\sqrt{\hat{s}}>1~TeV$) is for the Born cross
section and the lower one for the one-loop QCD corrected cross
section.

{\bf Figure \ref{fig:gg_s_cs}} The Born and one-loop electroweak
corrected cross sections for the \ggp subprocess as the functions
of c.m.s. energy ($\sqrt{\hat{s}}$) with $m_h=115$, $150$,
$200$~GeV. For each line type, the upper curve is for the Born
cross section and the lower one for the one-loop electroweak
corrected cross section.

\par
{\bf Figure \ref{fig:gg_s_rel_qcd}} The QCD one-loop relative
corrections as the functions of c.m.s. energy ($\sqrt{\hat{s}}$)
for the $\gamma\gamma \to t\bar{t}h^0$ subprocess with $m_h=115$,
$150$, $200$~GeV, respectively.

\par
{\bf Figure \ref{fig:gg_s_rel}} (a)The electroweak one-loop
relative corrections as the functions of c.m.s. energy
($\sqrt{\hat{s}}$) with $m_h=115$, $150$, $200$~GeV for the
$\gamma\gamma \to t\bar{t}h^0$ subprocess. (b)The QED and weak
one-loop relative corrections as the functions of c.m.s. energy
($\sqrt{\hat{s}}$) with $m_h=115$, $150$, $200$~GeV for the
$\gamma\gamma \to t\bar{t}h^0$ subprocess. For each line type, the
upper curve is for the QED corrected cross section and the lower
one for the weak corrected cross section.

\par
{\bf Figure \ref{fig:gg_h_rel_qcd}} The QCD one-loop relative
corrections as the functions of the mass of Higgs boson ($m_h$)
for the \ggp subprocess with $\sqrt{\hat{s}}=500$, $800$, $1000$,
$2000$~GeV, respectively.

\par
{\bf Figure \ref{fig:gg_h_rel}} The electroweak one-loop relative
corrections as the functions of the mass of Higgs boson ($m_h$)
for the \ggp subprocess with $\sqrt{\hat{s}}=500$, $800$, $1000$,
$2000$~GeV, respectively.

\par
{\bf Figure \ref{fig:ee_s_cs_qcd}} The Born and one-loop QCD
corrected cross sections for the \eep process as the functions of
c.m.s. energy ($\sqrt{s}$) with $m_h=115$, $150$, $200$~GeV,
respectively. For each line type, the upper curve(in the energy
region $\sqrt{s} > 1.4~TeV$) is for the Born cross section and the
lower one presents the one-loop corrected cross section.

\par
{\bf Figure \ref{fig:ee_s_cs}} The Born and one-loop electroweak
corrected cross sections for the \eep process as the functions of
c.m.s. energy ($\sqrt{s}$) with $m_h=115$, $150$, $200$~GeV,
respectively. For each line type, the upper curve is for the Born
cross section and the lower one presents the one-loop corrected
cross section.

\par
{\bf Figure \ref{fig:ee_s_rel_qcd}} The QCD one-loop relative
corrections to the \eep process as the functions of c.m.s. energy
($\sqrt{s}$) with $m_h=115$, $150$, $200$~GeV, respectively.

\par
{\bf Figure \ref{fig:ee_s_rel}} The electroweak one-loop relative
corrections to the \eep process as the functions of c.m.s. energy
($\sqrt{s}$) with $m_h=115$, $150$, $200$~GeV, respectively.

\par
{\bf Figure \ref{fig:ee_h_rel_qcd}} The QCD one-loop relative
corrections to the \eep process as the functions of Higgs boson
mass ($m_h$) with $\sqrt{s}=800$, $1000$, $2000$~GeV.

\par
{\bf Figure \ref{fig:ee_h_rel}} The electroweak one-loop relative
corrections as the functions of Higgs boson mass ($m_h$) with
$\sqrt{s}=800$, $1000$, $2000$~GeV, respectively.

\par
{\bf Figure \ref{fig:ee_mu_rel_qcd}} The QCD relative corrections
to the \eep process as the function of the QCD renormalization
scale $\mu$ , with $m_h = 115~GeV$ and $\sqrt{s} =
800,~1000,~2000~GeV$.

\end{document}